\documentclass{article}

\PassOptionsToPackage{numbers, compress}{natbib}

\usepackage[final]{neurips_2020}

\usepackage[utf8]{inputenc} 
\usepackage[T1]{fontenc}    
\usepackage{hyperref}       
\usepackage{url}            
\usepackage{booktabs}       
\usepackage{amsfonts}       
\usepackage{nicefrac}       
\usepackage{microtype}      
\usepackage{enumitem}
\usepackage{graphicx}
\usepackage{makecell}
\usepackage{algorithm}
\usepackage{algorithmic}
\usepackage{amsthm}
\usepackage{subcaption}

\usepackage{amsmath,amsfonts,bm}
\allowdisplaybreaks

\newcommand{\graph}{\mathcal{G}}

\newcommand{\MLP}{\mathrm{MLP}}

\newcommand{\set}[1]{\{ #1 \}}

\def\eqref#1{equation~\ref{#1}}

\def\1{\bm{1}}

\def\vb{{\bm{b}}}

\def\vh{{\bm{h}}}

\def\vw{{\bm{w}}}

\def\vz{{\bm{z}}}

\DeclareMathAlphabet{\mathsfit}{\encodingdefault}{\sfdefault}{m}{sl}
\SetMathAlphabet{\mathsfit}{bold}{\encodingdefault}{\sfdefault}{bx}{n}

\def\gD{{\mathcal{D}}}

\def\gL{{\mathcal{L}}}

\def\gV{{\mathcal{V}}}

\newcommand{\sigmoid}{\sigma}

\title{Discovering Synergistic Drug Combinations for COVID with Biological Bottleneck Models}

\author{%
  Wengong Jin $\quad$ Regina Barzilay $\quad$ Tommi Jaakkola \\
  CSAIL, Masssachusetts Institute of Technology \\
  \texttt{\{wengong,regina,tommi\}@csail.mit.edu}
}

\begin{document}

\maketitle
\begin{abstract}
Drug combinations play an important role in therapeutics due to its better efficacy and reduced toxicity. Recent approaches have applied machine learning to identify synergistic combinations for cancer, but they are not applicable to new diseases with limited combination data. Given that drug synergy is closely tied to biological targets, we propose a \emph{biological bottleneck} model that jointly learns drug-target interaction and synergy. The model consists of two parts: a drug-target interaction and target-disease association module. This design enables the model to \emph{explain} how a biological target affects drug synergy. By utilizing additional biological information, our model achieves 0.78 test AUC in drug synergy prediction using only 90 COVID drug combinations for training. We experimentally tested the model predictions in the U.S. National Center for Advancing Translational Sciences (NCATS) facilities and discovered two novel drug combinations (Remdesivir + Reserpine and Remdesivir + IQ-1S) with strong synergy in vitro.

\end{abstract}

\section{Introduction}

Combination therapies have shown to be more effective than single drugs in multiple diseases such as HIV and tuberculosis~\citep{tan2012systematic,yilancioglu2019design}. Synergistic combinations can improve both potency and efficacy, either achieving stronger therapeutic effects and/or decreasing dosage thereby reducing side-effects. In the times of current pandemic, finding a successful combination of approved molecules have an additional benefit over designing a de-novo molecule: time to clinical adoption. Approved drugs are typically commercially available and have well studied safety profiles. Taken in aggregate, these considerations motivate us to explore combination therapies for COVID antivirals.

Since exploring the space of combinations via high-throughput screening is prohibitively expensive as it involves combinatorial search, in-silico screening based on machine learning is an appealing alternative. In fact, a number of such methods have been reported in the literature~\cite{preuer2018deepsynergy,xia2018predicting}. These techniques have been shown effective when the model was provided with large amounts of training data capturing synergy of various combinations. Unfortunately, this requirement prevents us from utilizing these techniques for many diseases where such data is not available. 
Therefore, it is crucial to reduce data dependence to make combination algorithm applicable in multiple therapeutic contexts. 

In this paper, we present a novel algorithm for finding combinations that achieves this goal. Our main hypothesis is that by explicitly modeling interaction between compounds and the biological targets, we can significantly decrease dependence on combination training data. The proposed \emph{biological bottleneck model} has two components. The first component models drug-target interactions (DTI) predicting which targets are inhibited by a compound. It is trained on individual compounds since DTI information is readily available for multiple targets across multiple diseases. Our second component focuses on modeling target-disease association. 
It is a simple linear function which enables the model to explain how much a biological targets affects synergistic activity.

\begin{figure}
    \centering
    \includegraphics[width=\textwidth]{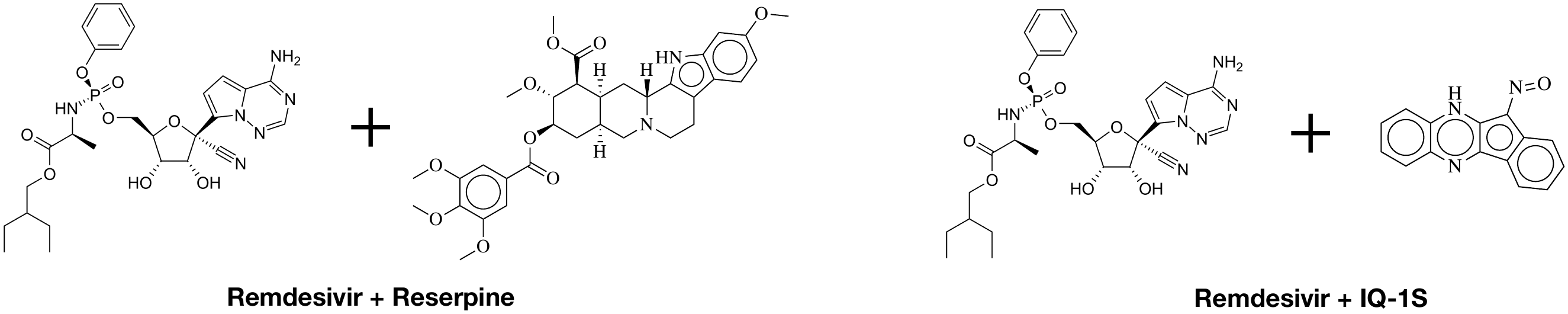}
    \caption{Our model discovered two novel drug combinations (Remdesivir + Reserpine and Remdesivir + IQ-1S) that show strong synergy in Vero E6 cells.}
    \label{fig:newdrugs}
\end{figure}

We develop our model using single agent and drug combination data from various sources. It incorporates known COVID biological targets~\citep{gordon2020sars} and their corresponding drug-target activity collected from ChEMBL~\citep{gaulton2017chembl}. With only 90 COVID drug combinations for training, our model achieves 0.78 test AUC on the SARS-CoV-2 combination screen from \citet{bobrowski2020discovery}. Moreover, incorporating known COVID targets yields 10\% relative increase in test accuracy. Lastly, we experimentally tested our model predictions in the NCATS facilities and discovered two novel drug combinations (Remdesivir + Reserpine and Remdesivir + IQ-1S) with strong synergy in Vero E6 cells (see Figure~\ref{fig:newdrugs}).

\section{Related Work}
Existing approach on drug synergy prediction can be roughly divided into two categories: 
\begin{itemize}[leftmargin=*,topsep=0pt,itemsep=0pt]
\item \emph{Supervised learning}: In this approach, a model is trained on combination data generated from high-throughput screens. For example, \citet{preuer2018deepsynergy} trained a deep neural network on a large-scale oncology screen~\cite{o2016unbiased} (23K training examples) to predict anti-cancer drug synergy. \citet{xia2018predicting} and \citet{sidorov2019predicting} trained deep neural networks to predict anti-cancer drug synergy on a larger dataset compiled by NCI-ALMANAC~\cite{holbeck2017national}, which contains around 300K training examples across 40 different cell lines. 

\item \emph{Biological networks}: Another category of drug synergy models are based on biological networks. Their assumption is that drugs with complementary mechanism of actions are more likely to be synergistic. For instance, \citet{cheng2019network} and \citet{zhou2020network} proposed to model synergy using distance metrics over drug-target interaction and protein-protein interaction networks. 
\end{itemize}
The major challenge of supervised approaches is the lack of combination data. For many diseases such as COVID and tuberculosis, the amount of drug combination data is very limited (less than 200)~\cite{bobrowski2020discovery,yilancioglu2019design}. Deep models are prone to over-fitting in this low-resource scenario. Moreover, as the number of pair-wise combinations grows quadratically with the number of drugs, the largest existing combination screen for cancer~\cite{holbeck2017national} only covers around 100 different drugs. This significantly limits the ability of trained models to generalize to new drugs outside of the training set.
On the other hand, while network-driven methods have a wider coverage over the chemical space, they cannot make predictions on new compounds outside of the network (i.e., drugs without target interaction data).

We propose a new method that combines the merit of both approaches while addressing their limitations. As drug interaction is often characterized by biological targets, our model is trained to predict both drug-target interaction and drug synergy. This enables us to make predictions on new compounds even if their drug-target interaction is unknown. This also addresses the data scarcity challenge since there are abundant drug-target interaction data available.

\section{Biological Bottleneck Models}

In this section, we describe our model architecture for drug combinations. A drug combination is called synergistic if its antiviral effect is greater than the sum of the individual effects. Drug synergy arises from various types of drug interaction. For example, two drugs can be synergistic when they interact with different sets of biological targets or pathways. Indeed, most of the anti-HIV drug combinations, such as Dolutegravir and Lamivudine, are drugs with different mechanisms of actions (i.e., interacting with different biological targets). To account for this inductive bias, it is crucial to model the interaction between drugs and biological targets in our model architecture.

Motivated by these observations, we propose to decompose our model into two parts: a drug-target interaction (DTI) module $\phi$ and a target-disease association module $f$. The DTI module predicts the biological targets activated by a given compound. The target-disease association module learns how a biological target is related to the disease. The vocabulary of biological targets are chosen by experts in advance. To introduce our method, we first describe how these two modules are used to predict antiviral activity of single compounds and then extend it to drug combinations. 



\begin{figure}
    \centering
    \includegraphics[width=\textwidth]{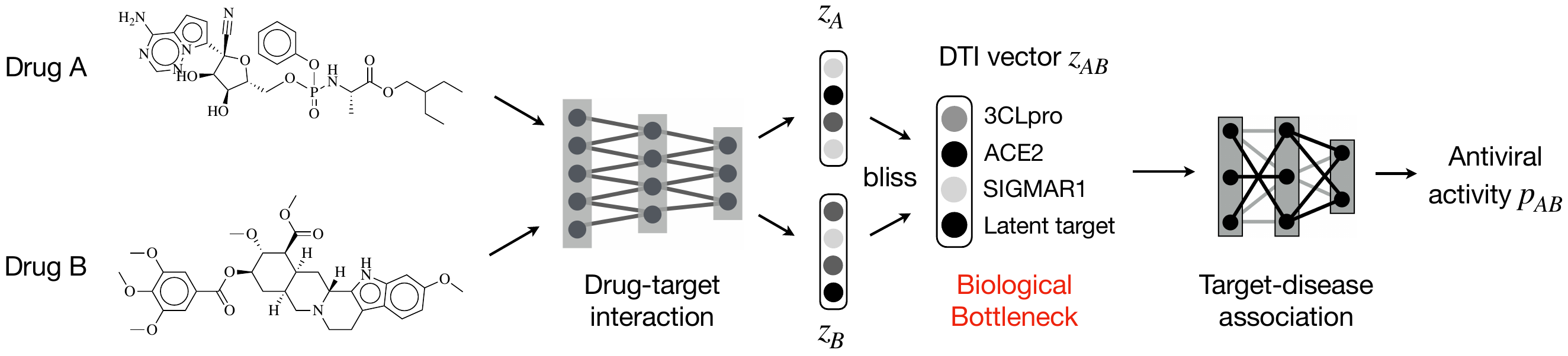}
    \caption{Our biological bottleneck model is composed of two modules: a drug-target interaction (DTI) and target-disease association module. The antiviral effect of a combination is predicted on their DTI vector $\vz_{AB}$, which is computed from the DTI vectors $\vz_A, \vz_B$ of each individual drug. The DTI vector characterizes the drug-target interaction profile of a given drug.}
    \label{fig:model}
\end{figure}

\subsection{Forward pass of single drugs}
We represent a drug as a graph $\graph$, whose nodes and edges represent atoms and  bonds. To predict the antiviral effect of a single drug $A$, our model needs to accomplish two tasks: 1) predict its interaction with biological targets $\mathcal{V}=\set{t_1,\cdots,t_m}$; 2) learn the relevance of each target $t_i$ to the disease. 

\textbf{Drug-target interaction } We parametrize the DTI module $\phi$ as a graph convolutional network (GCN)~\cite{duvenaud2015convolutional,gilmer2017neural}. The GCN translates a molecular graph $\graph_A$ into a continuous vector through directed message passing operations~\cite{yang2019analyzing}, which associate hidden vectors $\vh_v$ with each node $v$ and updates these vectors by passing messages $\vh_{uv}$ over edges $(u,v)$. The output of $\phi$ is a vector $\vz_A$ representing the biological targets activated by drug $A$:
\begin{equation}
    \vz_A = \sigmoid\big(\MLP\big(\sum_{v \in \graph_A}\nolimits \vh_v\big)\big) \qquad \set{\vh_v} = \mathrm{GCN}(\graph_A)
\end{equation}
where $\sigmoid(\cdot)$ is a sigmoid function and $\MLP$ is a two-layer feed-forward network. Each element $\vz_{A,k}$ represents the probability of drug $A$ inhibits target $t_k$. Each target $t_k$ is associated with a drug-target interaction dataset $\gD_k=\set{(X_i,y_i)}$, where $y_i=1$ if a drug $X_i$ is interacts with target $t_k$. We will train this module on the DTI dataset of all biological targets in the vocabulary.

\textbf{Target-disease association } We parametrize the target-disease association module $f$ as a simple linear layer $(\vw,\vb)$ due to its interpretability. As shown in Figure~\ref{fig:model}, our model predicts the antiviral activity of a drug $A$ as:
\begin{equation}
    p_A = f(\vz_A) = \sigmoid(\vw^\top \vz_A + \vb) 
\end{equation}

\subsection{Forward pass of drug combinations}

Synergy are often quantified under Bliss synergy score~\cite{bliss1939toxicity}. Suppose the individual antiviral effect of drugs $A$ and $B$ are $p_A, p_B$. The expected effect of combination $(A,B)$ is given as $e_{AB} = p_A + p_B - p_Ap_B$. A drug combination $(A,B)$ is synergistic if its observed effect $p_{AB} > e_{AB}$. Following this definition, we introduce a new Bliss layer to predict the synergistic effect of a drug combination $(A,B)$. Given two drugs and their predicted DTI vectors $\vz_A, \vz_B$, the Bliss layer computes the DTI vector $\vz_{AB}$ as
\begin{equation}
    \vz_{AB} = \vz_A + \vz_B - \vz_A \odot \vz_B
\end{equation}
where $\odot$ stands for element-wise multiplication. With this aggregation function, a drug combination will benefit most from \emph{complementary} targets. If only one drug is active to target $t_i$ (e.g., $\vz_{A,i}=1, \vz_{B,i}=0$), the combination $(A,B)$ is still active to $t_i$ ($\vz_{AB,i}=1$). In other words, the set of active targets for $(A,B)$ is the union of active targets of the two drugs.

Given a drug combination $(A,B)$, our model predicts its antiviral activity as:
\begin{equation}
    p_{AB}= f(\vz_{AB}) = \sigmoid(\vw^\top\vz_{AB} + \vb)
\end{equation}
Following the Bliss independence model, we predict the synergy score of a combination as $p_{AB} - e_{AB}$, where $e_{AB} = p_A + p_B - p_Ap_B$. Intuitively, a combination is more likely to be synergistic if they have complementary targets with high target-disease association score.

\subsection{Learning latent targets}
In order to predict synergy, it is important to incorporate all the relevant biological targets into our model. However, this is challenging for two reasons: First, most biological targets do not have drug-target interaction data and thus cannot be incorporated in our model. Second, current biological understanding of a disease may be incomplete. For instance, \citet{riva2020discovery} reported around 50 new biological targets related to COVID antiviral activity, but they are not reported in the previous work by \citet{gordon2020sars}.

To this end, we propose to include additional \emph{latent targets} in the bottleneck layer that are learned indirectly from single-agent and combination data. Specifically, we expand the dimension of $\vz_A$ to be greater than the total number of considered targets in $\gV$. The first $m$ entries in $\vz_A$ corresponds to the real biological targets and the other entries are latent targets. As we will show in the experiments, it is possible for us to interpret  new biological targets related to given diseases. 

\subsection{Training} 
Our training loss $\gL = \lambda_\mathrm{DTI} \ell_{\mathrm{DTI}} + \lambda_S \ell_S + \ell_C$ consists of three components. First, the drug-target interaction loss $\ell_{\mathrm{DTI}}$ enforces the DTI vector $\vz_A$ to be biologically meaningful. This is calculated for each target based on its DTI dataset $\gD_k = \set{(X_i,y_i^t)}$. The DTI module $\phi$ is trained to minimize:
\begin{equation}
    \ell_{\mathrm{DTI}} = \sum_k \sum_{(X_i,y_i^t) \in \gD_k} \ell(\vz_{X_i,k}, y_i^t) 
\end{equation}
Second, our model is trained on single-agent data $\gD_S = \set{(X_1,y_1^s),\cdots,(X_n,y_n^s)}$. Each molecule $x_i$ is labeled with its antiviral activity (active/inactive). Both modules $\phi,f$ are trained to minimize
    \begin{equation}
        \ell_S = \sum_{(X_i,y_i^s) \in \gD_S} \ell(f(\phi(X_i)), y_i^s) 
    \end{equation}
Lastly, the model is trained on drug combination data $\gD_C = \set{(A_i, B_i,y_i)}$. Each drug combination $(A_i,B_i)$ has a synergy label $y_i^c$, where $y_i^c=1$ means it is synergistic and $y_i^c=0$ additive or antagonistic. We train both modules $\phi,f$ to minimize
    \begin{equation}
        \ell_C = \sum_{(A_i, B_i,y_i^c) \in \gD_C} \ell(p_{A_i B_i} - e_{A_i B_i}, y_i^c)
    \end{equation}

\textbf{Multi-disease training } Since COVID is a new disease, its drug combination data is very limited. To address the low-resource challenge, we utilize additional drug synergy data from other viral diseases such as HIV. Specifically, we augment the model with HIV biological targets as well as HIV single-agent and drug combination data.  
The DTI module $\phi$ now outputs a DTI vector $\vz_A^d$ for each disease $d$. $\phi$ is shared across two diseases and trained to learn drug-target interaction for all diseases.
Since each disease operates on different targets, we create a target-disease association module $f^D$ for each disease. Let $\ell_{\mathrm{DTI}}^d, \ell_S^d, \ell_C^d$ be the losses for each disease $d \in \set{\text{COVID, HIV}}$. Our final training loss becomes
\begin{equation}
    \mathcal{L}_{\mathrm{multi}} = \sum_d \lambda_1 \ell_{\mathrm{DTI}}^d + \lambda_2 \ell_S^d + \ell_C^d 
\end{equation}

\section{Experiments}
\label{sec:experiment}

\begin{figure}
    \centering
    \includegraphics[width=0.9\textwidth]{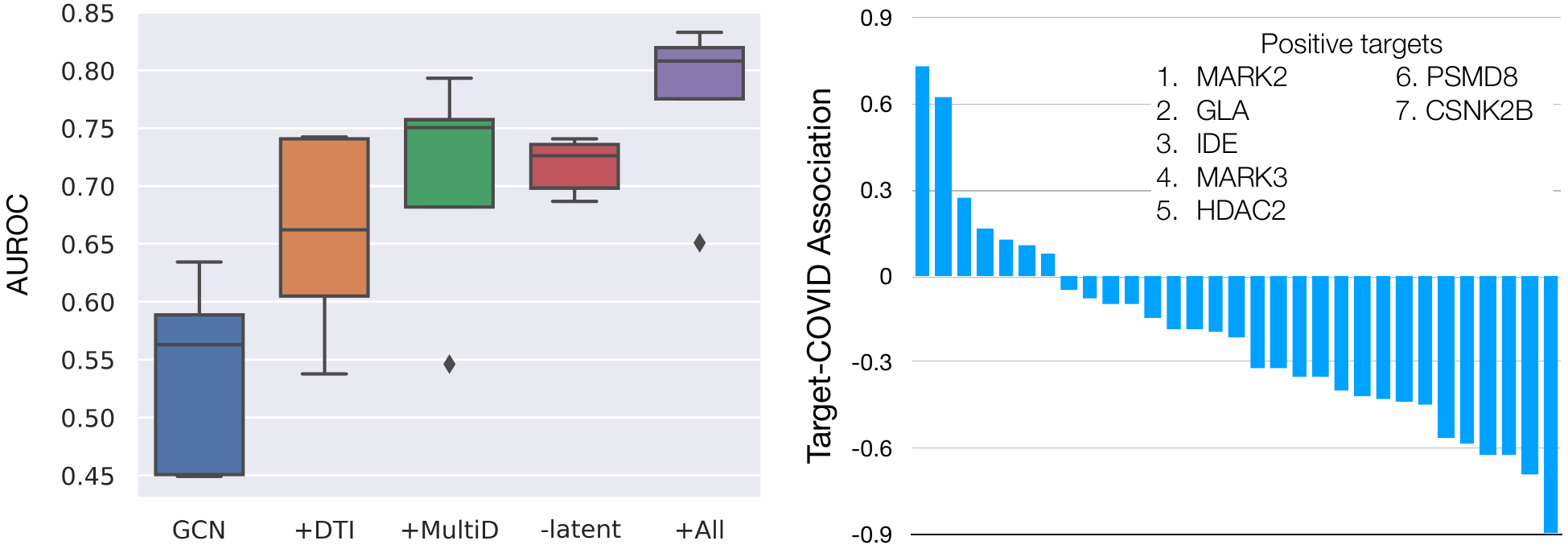}
    \caption{\emph{Left}: Results on SARS-CoV-2 combination test set. Our model (+all) outperforms all other baselines. \emph{Right}: Seven targets that positively contributes COVID drug synergy.}
    \label{fig:results}
\end{figure}

\textbf{SARS-CoV-2 Data } For SARS-CoV-2 infection, we consider three types of biological targets in our target vocabulary $\mathcal{V}=\set{t_1,\cdots,t_m}$:
\begin{itemize}[leftmargin=*,topsep=0pt,itemsep=0pt]
    \item \emph{Viral proteases}: Replication of SARS-CoV-2 virus requires the processing of two polyproteins by two virally encoded proteases: chymotrypsin-like protease (3CLpro) and papain-like protease (PLpro). Inhibitors that block either protease could inhibit viral replication. We have compiled 3CLpro enzymatic activity~\cite{ncats_3clpro} and PLpro inhibition~\cite{reframe_plpro} data made public by NCATS and ReframeDB.
    \item \emph{Viral entry proteins}: SARS-CoV-2 cell entry depends on angiotensin converting enzyme 2 (ACE2)~\cite{hoffmann2020sars}. Inhibiting ACE2 enzyme or the interaction between SARS-CoV-2 and ACE2 could block viral entry. To this end, we utilize ACE2 enzymatic activity~\cite{ncats_ace2} and Spike-ACE2 protein-protein interaction~\cite{ncats_spike_ace2} from NCATS.
    \item \emph{Host proteins}: \citet{gordon2020sars} identified 335 human proteins physically associated with SARS-CoV-2 viral proteins. Inhibitors for these proteins may also hinder viral replication. Among these proteins, we selected 31 proteins that have sufficient amount of drug-target interaction data in the ChEMBL database (i.e., both positive and negative interactions). 
\end{itemize}
The above drug-target interaction data contains around 20K compounds in total. Our training data for SARS-CoV-2 utilizes another two assays:
\begin{itemize}[leftmargin=*,topsep=0pt,itemsep=0pt]
    \item \emph{Single-agent Activity}: We use the NCATS CPE assay in VeroE6 cells~\citep{ncats_cpe}, which contains around 10K compounds and 320 hits with EC50 $\leq 10 \mu$M.
    \item \emph{Drug Combination}: NCATS performed two combination assays in VeroE6 cells, which contain 160 two-drug combinations~\cite{bobrowski2020discovery,ncats_matrix}. \citet{riva2020discovery} also analyzed synergy between Remdesivir and 20 active compounds identified from their high-throughput screen.
\end{itemize}

\textbf{HIV Data } The training data for HIV comes from the following assays:
\begin{itemize}[leftmargin=*,topsep=0pt,itemsep=0pt]
    \item \emph{Drug-target Interaction}: Existing anti-HIV drugs mainly target viral proteins (HIV-1 protease, integrase and reverse transcriptase) or host proteins involved in viral entry (CCR5, CXCR4 and CD4). We compiled DTI data for these six targets from ChEMBL. 
    \item \emph{Single-agent Activity}: NCI conducted an anti-HIV assay~\cite{nci_hiv} with 35K compounds, among which 309 compounds are active (EC50 $\leq 1 \mu$M).
    \item \emph{Drug Combination}: \citet{tan2012systematic} conducted high-throughput screen for HIV drug combinations. The dataset contains 114 two-drug combinations.
\end{itemize}

\textbf{Evaluation Protocol } Since our goal is to predict synergy against SARS-CoV-2, our validation and test set only consist of SARS-CoV-2 combinations.  All the drug-target interaction, single-drug activity and HIV data are used for training only.
Our validation set contains 20 combinations from \citet{riva2020discovery} and test set contains 72 combinations from \citet{bobrowski2020discovery}. The training set contains 90 SARS-CoV-2 combinations from \cite{ncats_matrix}, where we remove combinations that appear in both the training and test set. 

\textbf{Hyperparameters } For DTI module $\phi$, we adopt default hyperparameters from \citet{yang2019analyzing}, with hidden dimension 300 and three message passing iterations. We set the dimension of DTI vector $|\vz|=100$, with 42 real biological targets (SARS-CoV-2 and HIV) and 58 latent targets, so that the number of real and latent targets are roughly equal. We set $\lambda_1=10, \lambda_2=0.1$ for our final model.

\subsection{Results}
To show the effectiveness of different components, we compare with the following baselines:
\begin{itemize}[leftmargin=*,topsep=0pt,itemsep=0pt]
    \item A GCN trained only on SARS-CoV-2 single-agent and combination data ($\lambda_\mathrm{DTI}=0, |\vz|=100$).
    \item +DTI: A GCN trained only on SARS-CoV-2 single-agent, combination as well as drug-target interaction data ($\lambda_\mathrm{DTI}=10, |\vz|=100$).
    \item +MultiD: A GCN trained on both SARS-CoV-2 and HIV data (single-agent + combination), but without drug-target interaction data ($\lambda_\mathrm{DTI}=0, |\vz|=100$).
    \item +All,-latent: A GCN trained on both SARS-CoV-2 and HIV data (single-agent + combination + drug-target interaction), but the latent targets are removed ($\lambda_\mathrm{DTI}=10,|\vz|=42$).
    \item +All: A GCN trained on both SARS-CoV-2 and HIV data (single-agent + combination + drug-target interaction) ($\lambda_\mathrm{DTI}=10,|\vz|=100$).
\end{itemize}
Our results are shown in Figure~\ref{fig:results}. As expected, the GCN baseline performs poorly, with $0.537 \pm 0.075$ AUC. Adding drug-target interaction data (+DTI) improves the AUC to $0.658 \pm 0.079$. Adding HIV data (+MultiD) improves the AUC to $0.706 \pm 0.088$. 
Our final model, trained with both HIV and drug-target interaction (+All), achieves the best AUC of $0.777 \pm 0.066$. This validates the advantage of adding drug-target interaction data and multi-disease training. Note that if we remove the latent targets (+All,-latent), the performance decreases to $0.718 \pm 0.021$. This also shows the importance of using latent targets to complement missing biological information.

In Figure~\ref{fig:results}, we report the learned target-disease association score for all COVID targets. There are seven targets positively correlated with COVID antiviral activity. According to \citet{gordon2020sars}, these targets interact with SARS-CoV-2 Orf9b, Nsp14, Nsp5 and N viral proteins. The wide range of host-viral protein interaction indicates that drug synergy arises from different modes of action.

\begin{figure}[t]
    \centering
    \begin{subfigure}[t]{0.24\textwidth}
    \includegraphics[width=\textwidth]{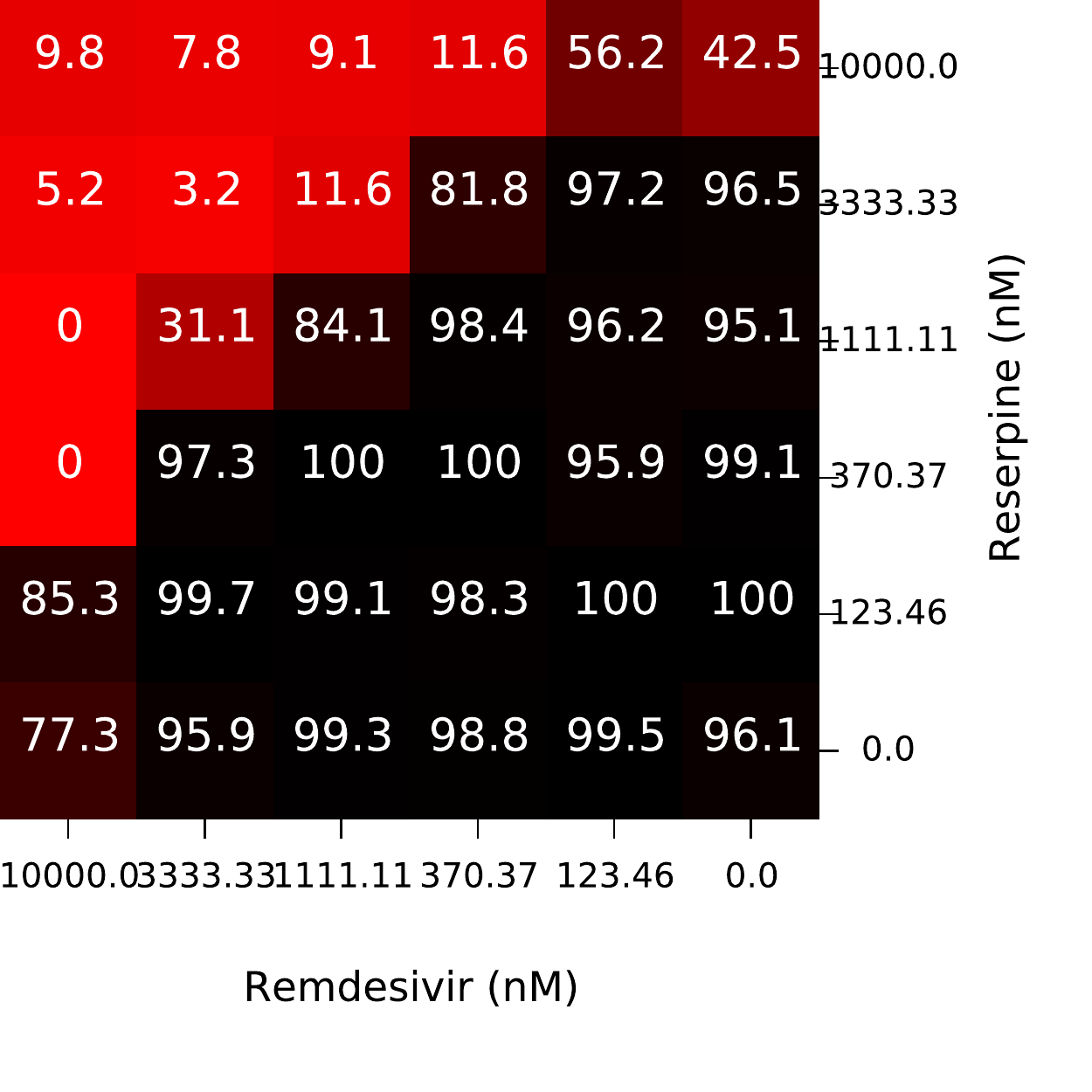}
    \end{subfigure}
    \begin{subfigure}[t]{0.24\textwidth}
    \includegraphics[width=\textwidth]{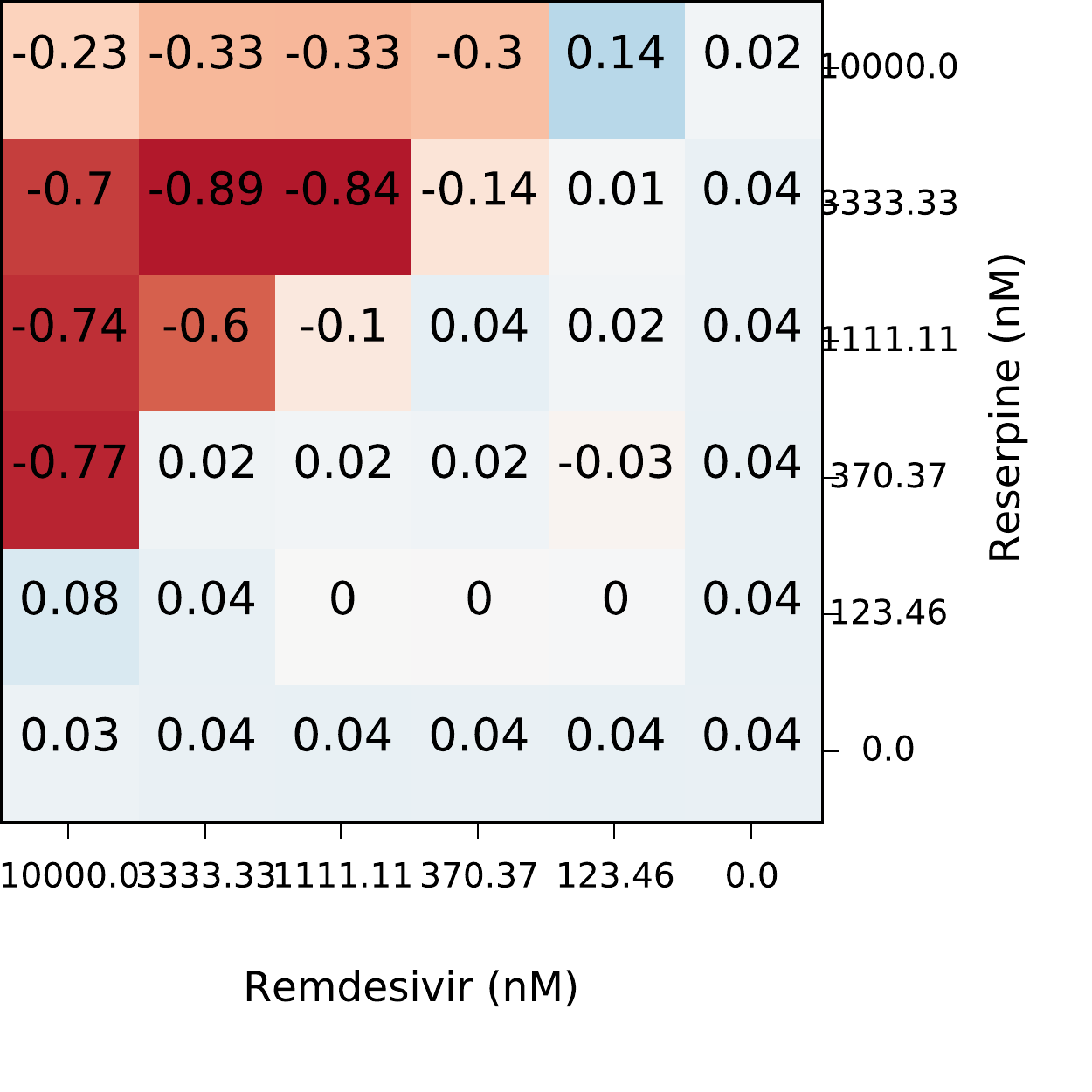}
    \end{subfigure}
    \begin{subfigure}[t]{0.24\textwidth}
    \includegraphics[width=\textwidth]{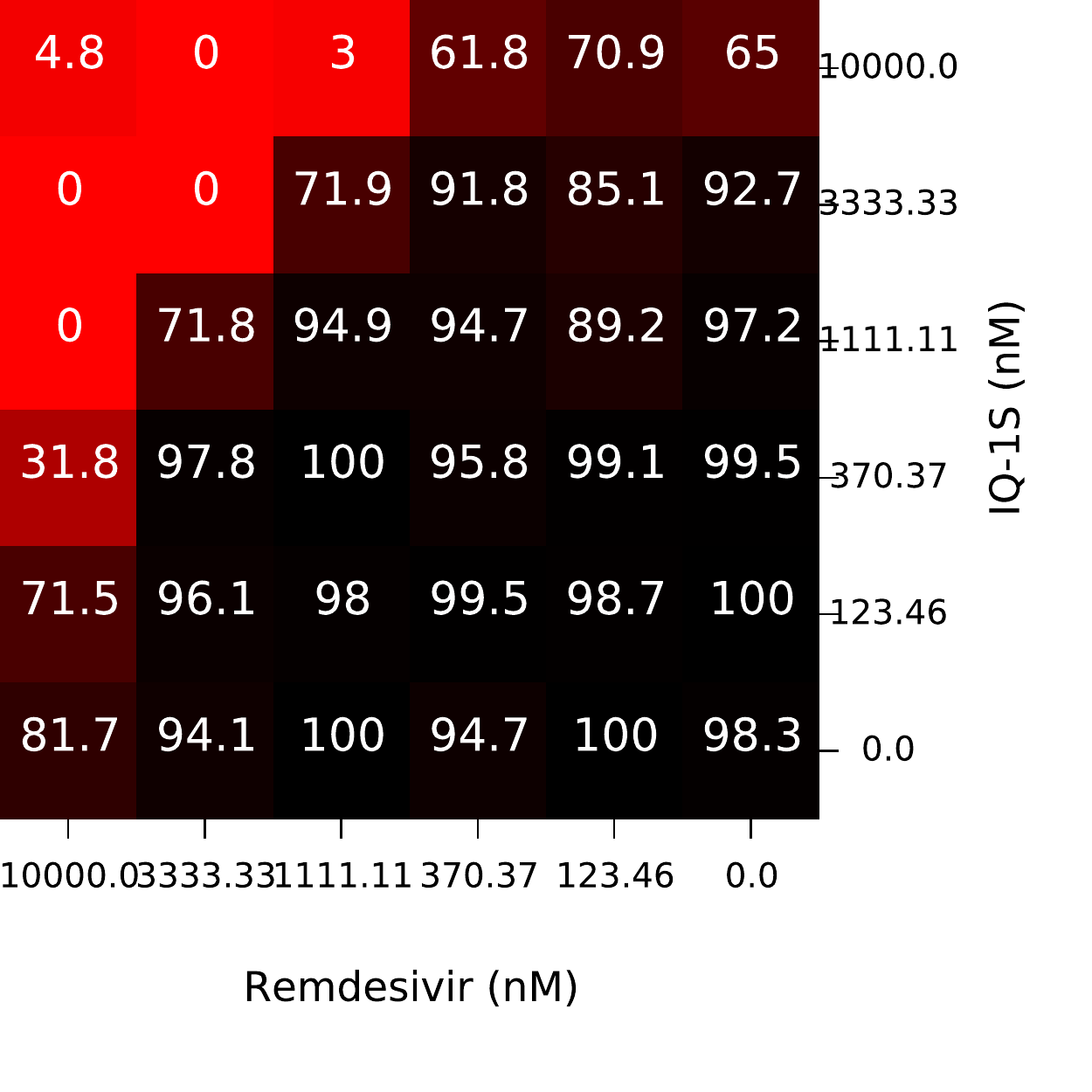}
    \end{subfigure}
    \begin{subfigure}[t]{0.24\textwidth}
    \includegraphics[width=\textwidth]{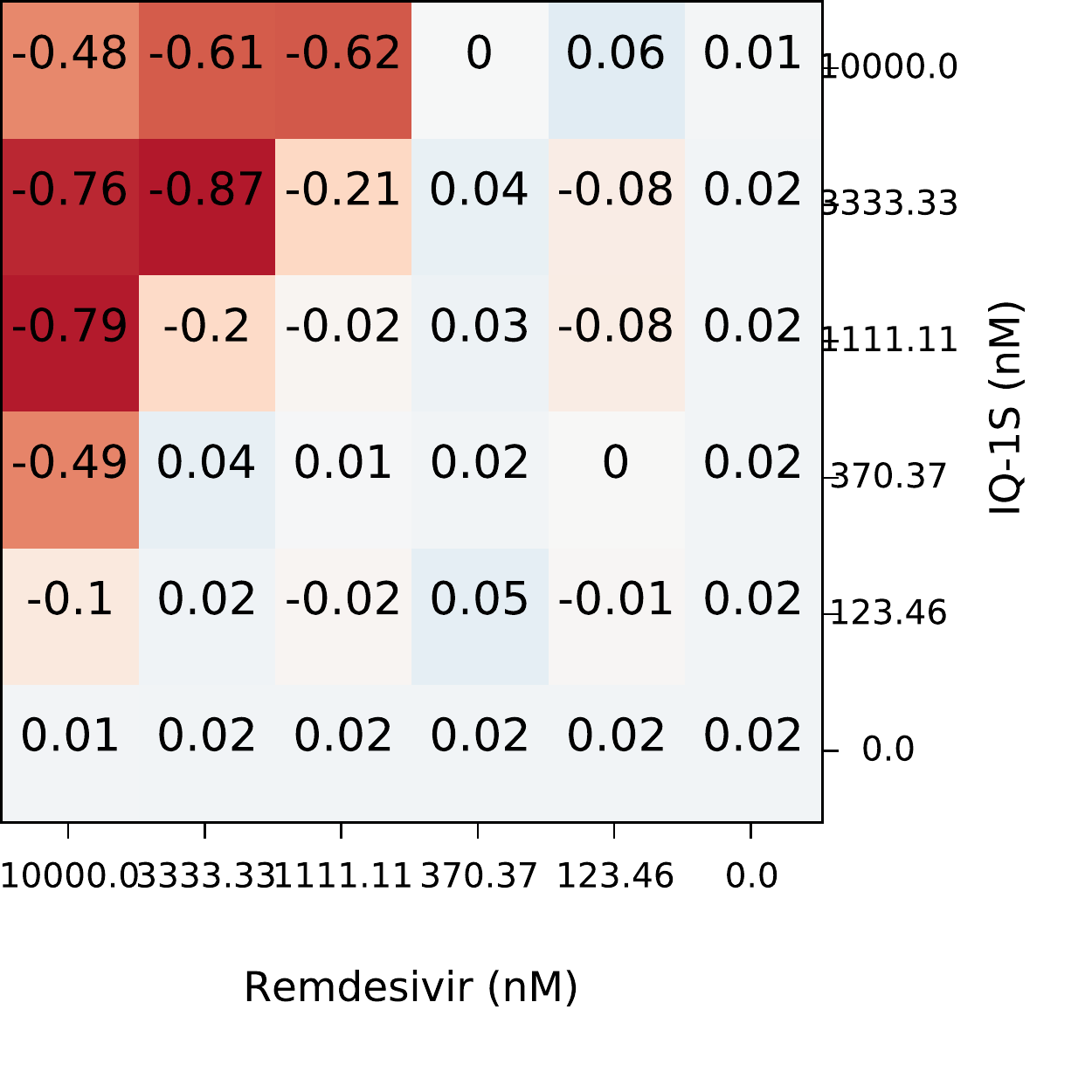}
    \end{subfigure}
    \caption{Two new drug combinations are discovered through our model: Remdesivir + Reserpine and Remdesivir + IQ-1S. Figures on the left show the dose response matrix and bliss synergy matrix for Remdesivir + Reserpine. Figures on the right show the same information for Remdesivir + IQ-1S.}
    \label{fig:matrix}
\end{figure}

\subsection{Discovering new synergistic combinations}
We applied our best model to predict synergy of drug combinations in the NCATS compound library. Given limited experimental resources, we only evaluated pairwise combinations between potent drugs with single-agent IC50 less than 30uM. This gave us around 11600 drug combinations ranked according to predicted synergy score with the highest scores. We selected the top 30 candidates and experimentally tested them in NCATS facilities.\footnote{The detailed protocol for viral cytopathic effect assay (CPE) and toxicity counter assays are available from NCATS OpenData portal: \url{https://opendata.ncats.nih.gov/covid19/assay?aid=14}}
We successfully discovered two new drug combinations (Remdesivir + Reserpine and Remdesivir + IQ-1S) with strong synergy in Vero E6 cells. The dose response and bliss synergy matrix are shown in Figure~\ref{fig:matrix}.

\section{Conclusion}
In this paper, we present a biological bottleneck model for predicting COVID drug synergy. The model consists of two components: a drug-target interaction module and target-disease association module. Our model requires much less combination data since it utilizes additional biological information. The proposed model is general and applicable to other diseases. 

\nocite{*}
\bibliography{main}
\bibliographystyle{plainnat}

\end{document}